# Studies of THGEM-based detector at low-pressure Hydrogen/Deuterium, for AT-TPC applications

Marco Cortesi,[*] John Yurkon, Wolfgang Mittig, Daniel Bazin, Saul Beceiro-Novo and Andreas Stolz

*National Superconducting Cyclotron Laboratory, Michigan State University,
East Lansing, Michigan 48824, USA*

   *E-mail*: Cortesi@nscl.msu.edu

ABSTRACT: We study the performance of single- and double- THick Gas Electron Multiplier (THGEM) detectors in pure Hydrogen ($H_2$) and Deuterium ($D_2$) at low pressures, in the range of 100-450 torr. The effect of the pressure on the maximum achievable gain, ion-back flow and long-term gain stability are investigated for single and double cascade detectors. In particular, it was found that maximum achievable gains above $10^4$, from single-photoelectrons avalanche, can be achieved for pressures of 200 torr and above; for lower pressure the gains are limited by avalanche-induced secondary effects to a values of around $10^3$. The results of this work are relevant in the field of avalanche mechanism in low-pressure, low-mass noble gases, in particular for applications of THGEM end-cap readout for active-target Time Projection Chambers (TPC) in the field of nuclear physics and nuclear astrophysics.

---

[*] Corresponding author.

# Contents

## 1. Introduction

Time Projection Chambers (TPC) are widely used in nuclear and particle physics since they were first introduced by Nygren in the late 1970s [1]. In its basic design, a TPC consists of a gas-filled detection volume in an electric drift field, terminated at one end with an electron avalanche multiplier structure coupled to a high-granularity position-sensitive pad readout. In the field of modern nuclear physics and astrophysics, TPCs feature an active target mode (AT-TPC), where the filling gas is the electron avalanche medium and the target at the same time. AT-TPCs are used to study reactions in inverse kinematics induced by low-energy rare isotope beams, such as fusion, isobaric analog states, cluster structure of light nuclei and transfer reactions, without significant loss in resolution due to the thickness of the target [2], [3]. The identity of the gas used to fill the detector will depend upon the requirements of the particular physics case of interest. For instance, for proton scattering or for (p,x)-type single or multi-nucleon transfer reactions, hydrogen ($H_2$) as a proton target; deuterium ($D_2$) as deuteron target; helium (He) as an alpha target. Among many advantages, AT-TPCs allow for large dynamic ranges by adjusting the pressure of the filling gas target, and provide full solid angle acceptance. As the reactions being studied are typically very low in cross section, particularly in the nuclear astrophysics domain, operation of AT-TPCs in pure gases will allow to optimize the reaction yield and to suppress background from ambiguous underlying reaction mechanisms.

    One of the most recent improvement of the next generation AT-TPCs is the upgrade of the electron amplification technology from traditional wire-based detectors to micro-pattern gaseous detectors (MPGD) [4], [5]. On our previous work [6], we have reported the study of a Thick Gaseous Electron Multiplier (THGEM) detectors in pure, low-pressure Helium. Stable operational conditions and maximum detector gains of $10^4$-$10^7$ have been achieved at a pressure ranging from 100 torr up to 760 torr, with an estimated energy resolution of 2.4% (FWHM) for 5.5 MeV alpha particle tracks. In this work we report the characterization of a THGEM in pure hydrogen ($H_2$) and deuterium ($D_2$) for pressures in the range of 100-450 torr, including the study of the effective gain curve, ion-back flow and long-term gain stability.

## 2. Experimental setup and methodology

The experimental setup is conceptually similar to the one used in our previous study [6] and is schematically depicted in figure 1. It is composed of a 10x10 cm$^2$ THGEMs assembly, in single WELL- or double-THGEM cascade configurations, mounted in a single volume detector vessel of around 40 litres. In the case of two cascade THGEM setup, the last electrode was mounted in the WELL configuration [7]. The inner gas volume is sealed with an O-ring and evacuated to a residual pressure of a few mtorr for several hours before gas filling and data taking.

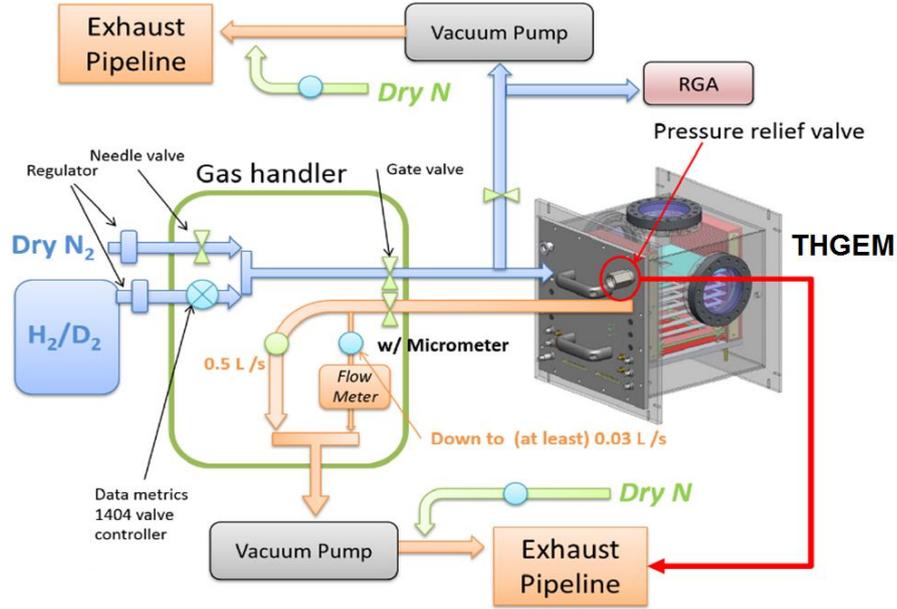

**Figure 1. Schematic drawing of the experimental setup.**

For safety reasons, the detector was operated in Hydrogen/Deuterium with pressures ranging from 100 to 450 torr with no steady flow of gas during operation. In order to avoid variations of the detector's effective gain due to a change of the gas impurities levels, data taking was performed several hours after the detector vessel was filled at the desired pressure. Under these conditions, gas impurities released by outgassing from assembly materials and from the gas system components, mostly nitrogen, are believed to reach a steady-state level below 0.5% (see for example [8]).

The THGEM used in this work and presented in the previous, related article [6], is based on avalanche multiplication in independent holes, arranged in a compact hexagonal array. The holes, with a 0.5 mm diameter are mechanically drilled in a copper-cladded double-sided 0.6 mm thick FR-4 plate, then the copper is etched to produce a 0.1 mm clearance from the holes' rim. The THGEM effective area is 10x10 cm$^2$. The distance between the THGEMs in the double cascade configuration is 2 mm. The bare-copper, top-surface of the first cascade THGEM was illuminated with a UV Hg(Ar)-lamp though a quartz window, resulting in a constant flux of emitted photoelectrons. Those electrons are focused and multiplied in the THGEM hole by the strong dipole electric field established within the holes, as a result of the potential difference between the two THGEM surfaces or between the top-surface of the WELL-THGEM and the readout anode. In the case of a double cascade assembly, an extraction field in the gap between the THGEM elements is responsible for an efficient transfer of the charges onto the second multiplying electrode. The bias-voltages of the various detector electrodes were supplied by individual power supplies via 10 MΩ resistors.

The effective gain was computed in current mode, by calculating the ratio of the current collected on the anode after avalanche multiplication to the photoelectron current extracted from the illuminated THGEM surface. The photoelectron current was determined in a dedicated set of measurements, in which a reverse electric field was applied on the drift gap and the photocurrent was recorded on the cathode mesh [6]. Current measurements were performed using a precision electrometer, either Keithley Model 614 or Keithley Model 480 [9].

Finally, the ion-back flow (IBF), defined as the fraction of avalanche-ions that drift back to the cathode, was measured as the ratio

$$\text{IBF} = \frac{I_+}{I_-} \qquad \text{Eq. 1}$$

where $I_+$ is the total positive (ion) current recorded on the cathode-mesh (measured by an electrometer connected to ground), while $I_-$ is the electron avalanche current measured as a voltage drop on the 10 MΩ bias-resistors of the WELL-THGEM anode. Figure 2 illustrates the schematic drawing of the experimental setup used for assessing the ion-back flow for a double-THGEM configuration.

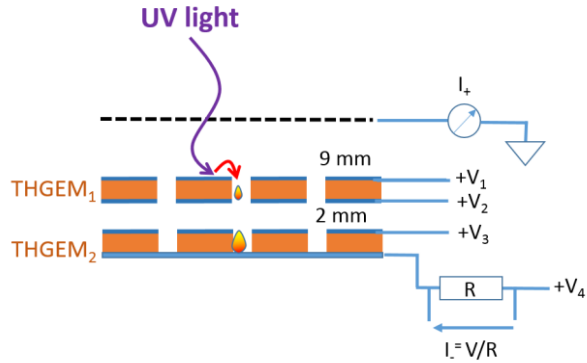

**Figure 2. Schematic drawing of the setup for ion-feedback measurements with a double-THGEM detector.**

## 3. Results

### 3.1 Effective gain

Figure 3 depicts the single-THGEM effective gain curve measured in Hydrogen ($H_2$ curve – open symbols) and in Deuterium ($D_2$ curve – full symbols), for pressures ranging from 100 to 450 torr. As a result of same electronic states of $H_2$ and $D_2$, no significant differences are observed in the effective gain curves of the two gases for any given pressure. The small discrepancies at low gains arise from the different current resolution of the pico-ammeters used for the two measurements.

Notice that diatomic molecules, like $H_2$ and $D_2$, are characterized by many rotational and vibration modes that can be excited by electron impact. These processes have relatively high cross sections that extend beyond the energy threshold of the ionization process. Under these conditions, an extremely high electric field strength is necessary to allow electrons to gain enough average energy to ionize the gas and initiate a substantial gas avalanche multiplication. For instance, figures 3 illustrates the cross sections for various electron-impact collision processes (elastic and inelastic) for $H_2$ (figure 3a) and He (figure 3b), as function of the electron

– 3 –

kinetic energy. While He is characterized by a few elastic and excitation processes, with a cross section that becomes comparable to the ionization cross section already at electron energy above 100 eV, the $H_2$ molecules have many elastic, rotational (ROT), vibrational (VIB), excitational (EXC) modes dominating over the ionization process for a wide range of electron energy spectrum. As a result compared to He and other gas-filling mixtures (including pure noble gases such as He, Ne or Ar), gas electrons avalanche multiplication in pure $H_2$ or $D_2$ requires several times larger operational voltages, with a significant probability of micro-discharges due to field emission and other secondary processes resulting in a limited maximum achievable gain.

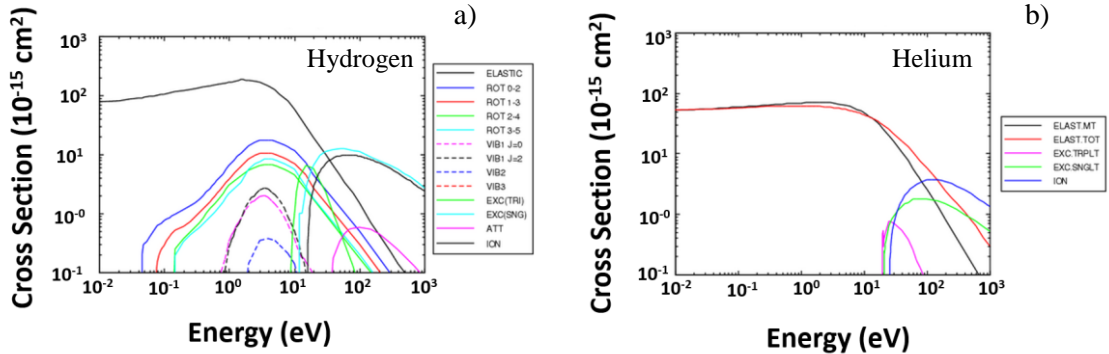

**Figure 3. Various cross section as function of the electron kinetic energy for $H_2$ (part a) and He (part b); graphs taken from [10].**

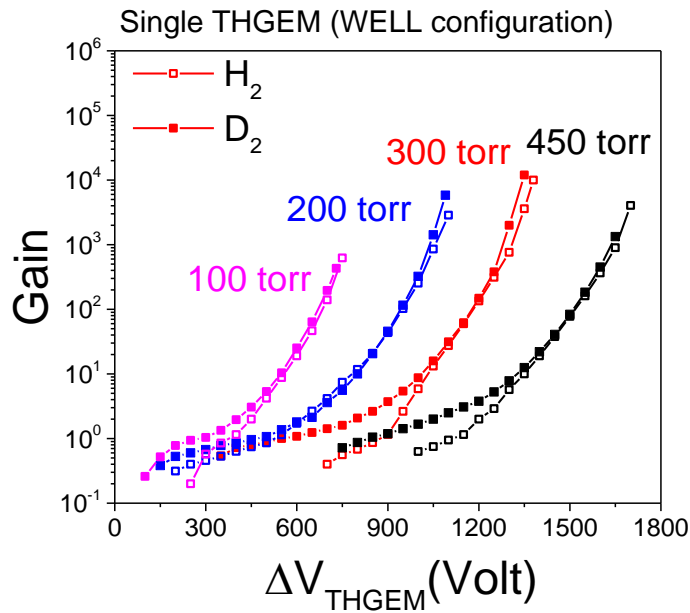

**Figure 4. Effective Gain of Single-THGEM (WELL configuration) in $H_2$ and $D_2$.**

As shown in figure 4, in $H_2$ and $D_2$ the single-THGEM detector allows maximum achievable gains up to $10^4$, for a pressure of 200 torr and above. For lower pressures, photo-mediated secondary effects becomes relevant at lower field strength, causing instabilities that prevent reaching high electron avalanche multiplication. At 100 torr the effective gain is limited to $10^3$.

A comparison of the effective gain for single and double-THGEM configurations, in pure $H_2$ as function of the gas pressure, is depicted in figure 5. Because of a large transversal diffusion of the electrons during the transfer from the first THGEM element to the second one, a larger effective gain is possible in a double-THGEM detector, while a lower voltage is applied to each single multiplier. The extremely high voltage needed for operation at high pressure, particularly the high bias needed to the first THGEM element, triggers self-sustained Townsend discharge, making the operation at high voltage quite difficult. As a result, the maximum achievable effective gain achieved at 450 torr is several times lower than the one reached at lower pressure.

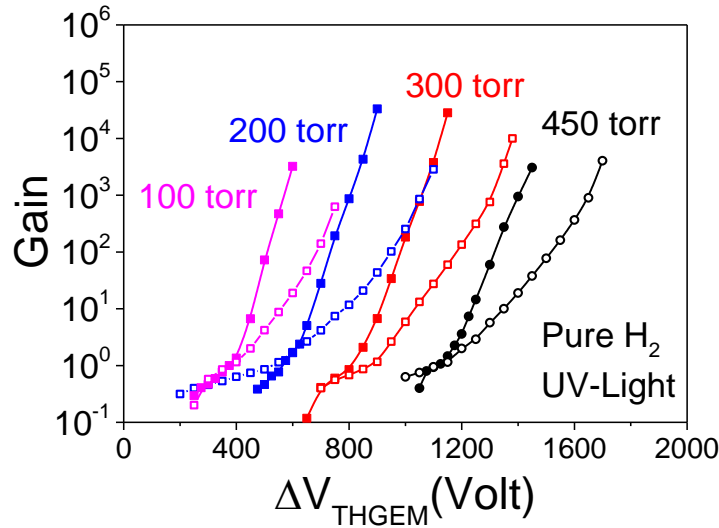

**Figure 5.** Comparison of the effective Gain measured with single- (open symbols) and double- (full symbols) THGEM detector. $\Delta V_{THGEM}$ is the voltage different applied to each single THGEM elements. An electric field of 0.5 kV/cm was applied between the THGEM elements (transfer field).

**3.2 Long-term gain Stability**

The long-term stability of the THGEM-based detector has been investigated using the same experimental methodology reported in [6]. According to our previous work, a large variation of the effective gain has been observed as a function of the chemical composition of the filling gas; in particular stabilization of the level of the residual impurities in a low-pressure environment was the main cause of a significant gain variation, occurring just after the detector vessel pressurization. Minor effects on the gain stabilization were observed as a consequence of the radiation-induced charging-up and of the polarization of the insulator substrate (FR-4); both processes are contingent exclusively on the voltage difference applied across the THGEM element.

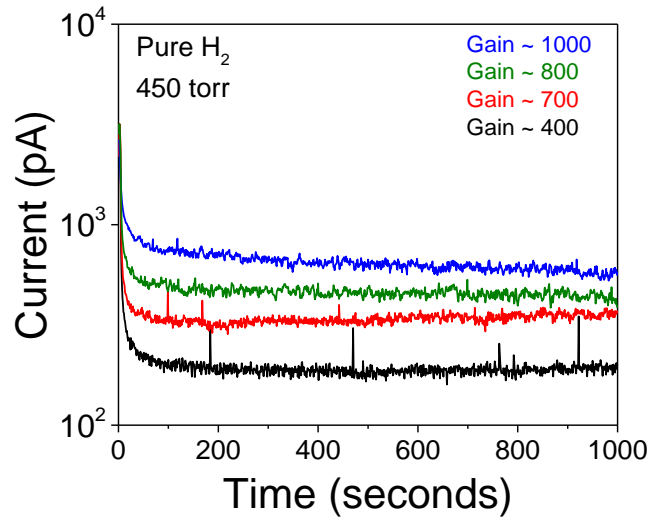

**Figure 6. Variation of the current recorded on the anode of the double THGEM setup as consequence of the effective gain variation due to charging up and/or polarization effects.**

With the intention of studying charging-up effects on the long-term stability of the effective gain independently from the gas impurities variations, the detector vessel was flushed at constant flow rate of $H_2/D_2$ for few minutes and then sealed for several hours, prior to any data taking. A two cascade THGEM assembly was tested under different conditions of bias-voltage and pressure, while illuminated for about 1000 seconds. The transfer field was kept constant at 0.5 kV/cm. The variation of the current recorded at the anode as a function of time is shown in figure 6, for different detector gain in $H_2$ at a pressure of 450 torr. After all the elements of the detector were biased, the UV lamp was switched on (t=0) initiating the data taking with a current sampling rate of 10 Hz. As illustrated by figure 6, a fast stabilization of the effective gain occurs within a characteristic time scale of a few tens of seconds, independently of the bias applied to the THGEM and thus to the detector gain. On a long term scale (several hundreds of second), we observed a much smaller gain variation of around a few percent. For instance, in the case of gain = 1000 and gain = 800 in figure 6, the current collected on the anode decreased by 12% and 5% respectively, over several hundreds of seconds. On the contrary, at gain = 700, we observed a current increase of 7%. These effects may be correlated to the "history" of the electrodes, i.e. being the first, second or third measurement cycle, or may depend on the condition and period of the previous polarization.

Notice that the few peaks discernable at low gain are probably due fluctuations of the pico-ammeter reading and they disappear when the measured current is significantly higher than its nominal current resolution.

### 3.3 Ion-back Flow

As a consequence of the gas amplification taking place in the THGEM holes, ions created during the avalanche processes may migrate towards the cathode and introduce electric field distortion in the drift volume. A severe ion-back flow could affect the performance and localization accuracy of the TPC and could trigger instabilities due to cathode excitation effects and electrons jet-induced breakdown [11]. Multi hole-type structures in cascade arrangements have proven to provide a natural capability to suppress ions from the avalanche in standard condition (atmospheric pressure) and in standard gas mixtures – see for example [12], [13] and references therein. However, no data are available at low pressure and for pure $H_2$ and $D_2$.

With the double THGEM detector setup depicted in figure 2, an extensive characterization of the ion-back flow, as function of the drift and transfer fields settings and of the effective gain, has been carried out in pure $H_2$ at a pressure of 450 torr. The range of the drift field considered in this study was however limited by the extreme operational conditions when high bias-voltage is applied to the detector elements. In figure 7 part a), the IBF is shown as a function of detector gain for three different drift field strengths (from 125, 250 and 375 V/cm); the electric field between the THGEM electrodes was kept constant at a value of 500 V/cm. In figure 7 part b), the IBF is depicted as function of the gain for three different transfer field, while the drift is kept constant at a value of 250 V/cm.

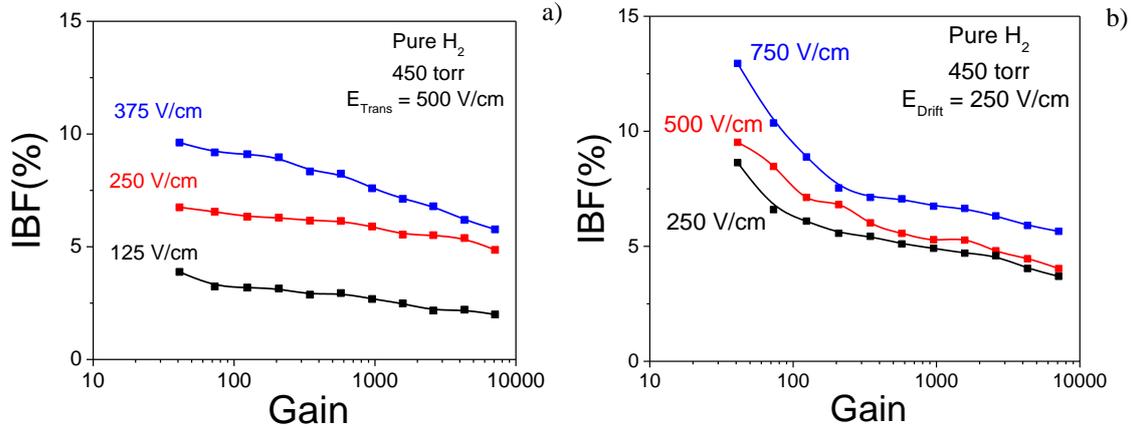

**Figure 7.** Ion back-flow (IBF%) as function of the detector gain, for different drift field (125, 250 and 375 V/cm), and for different transfer field (250, 500 and 750 V/cm), depicted respectively in part a) and part b). All the measurements were performed in $H_2$ at a pressure of 450 torr.

As shown in figures 7, the fraction of the ion trap on the THGEM electrode are very sensitive to the electric field geometry, which depend upon the field strength between the THGEM elements, on the field strength on the drift volume and on the avalanche field across the THGEM holes (namely the effective gain). A substantial fraction of ions is trapped on the THGEM electrodes when the field strength across the THGEM holes is much stronger than the transfer/drift field, with a substantial decrease of the IBF below 5% level when effective gain is above $10^3$.

## 4. Conclusion and Discussion

Operation of an AT-TPC detectors in low-pressure, low-mass gases for the study of inverse kinematic reactions, is a crucial factor for optimizing the reaction yield and for a unique and simpler disentanglement of the underlying reaction mechanisms. In this context, high efficiency will be important for the new frontier of low-intensity exotic-nuclear beam experiments in the astrophysics domain, where the reactions of interest have generally very low cross sections.

The performance of a THGEM-based detector in low-pressure, pure $H_2$ and $D_2$ has been systematically investigated. The main advantages of THGEM, compared to other MPGD structures such as standard GEM [14] or MICROMEGAS [15], is the extended thickness of the multiplication region within the THGEM holes, several times larger than the mean free path of the avalanche electrons, allowing to attain high electron-avalanche multiplications even at low pressure. In addition, photon-mediated secondary effects are considerably reduced due to the strong confinement of the avalanche within the THGEM holes, resulting in a more stable operation condition.

The operation and the amplification gains measured in extremely pure noble gases is considerably limited by the large amount of unquenched photons produced during the avalanche process. These avalanche photons generate secondary avalanches, feedback loop or photoionization of the gas that trigger the proportional avalanche to streamer transition and finally to the gas breakdown (discharge). On the other hand, small impurity admixtures to the noble gas, from the natural outgassing of the detector components, resulted in high reachable gains due to a substantial quenching effect of $N_2$ molecules ($N_2$ acts as wavelength shifter suppressing the UV component of the avalanche light).

Effective gains of the order of $10^4$, measured from single-electron avalanche, have been achieved for pressure ranging from 200 to 400 torr for single and double-cascade detector. Maximum effective gains of the order of $10^3$ were achieved with pressure lower than 200 torr. The gas amplification gains were mainly limited by the extremely high bias voltage necessary to reach electron avalanche multiplication in $H_2$ and $D_2$, causing a higher probability of sporadic discharge due to field emissions or secondary effects.

It has been observed that after impurities in the filling gas reached a steady concentration level, the detector provided a relative constant gas gain. Equilibrium between the main gas component (i.e. $H_2$ or $D_2$) and the impurities, outgassing from various detector elements, is attained after several hours. Moreover, typical gain variations due to charging-up and polarization effects of the THGEM insulator substrate are characterized by a fast decay-time (few tens of seconds). The Ion-back flow measured with a double-cascade THGEM detector at low pressure in $H_2$ and $D_2$ is of the order of 5-10%, depending on the drift/transfer field setup. Larger maximum gain and better ion-feedback suppression, leading to a lower spark probability, may be achieved with three THGEMs in cascade or combined THGEM and Micromegas detectors in a hybrid configuration.